\documentstyle[prd,aps,preprint,psfig]{revtex}
\begin{document}
\newcommand{\be}{\begin{equation}}
\newcommand{\ee}{\end{equation}}
\newcommand{\bea}{\begin{eqnarray}}
\newcommand{\eea}{\end{eqnarray}}
\newcommand{\vp}{\varphi}
\newcommand{\pd}{{\partial}}

\newcommand{\srr}{1/r^{2}}

\draft


\hsize\textwidth\columnwidth\hsize\csname @twocolumnfalse\endcsname

\draft

\title{Anisotropic Stars: Exact Solutions}

\author{Krsna Dev and Marcelo Gleiser}

\address{ Department of Physics and Astronomy, Dartmouth College,
Hanover, NH 03755 USA.}


\maketitle

\begin{abstract}
\noindent We study the effects of anisotropic pressure on the properties
of spherically symmetric, gravitationally bound objects. We consider 
the full general-relativistic treatment of this problem and obtain exact 
solutions for various forms of the equation of state connecting the 
radial and tangential pressures. 
It is  shown that pressure anisotropy can have 
significant effects on the structure and properties of stellar objects. In 
particular, the maximum value of $2M/R$ can approach unity ($2M/R< 8/9$ for
isotropic objects) and the surface redshift
can be arbitrarily large.     
\end{abstract}


\setlength{\baselineskip}{8.25mm}

\narrowtext
\section{Introduction}
\noindent A common assumption in the study of stellar structure 
and evolution is  
that the interior of a star can be modeled as a perfect fluid 
\cite{CLAYTON,KIPP}.  This
perfect fluid model  necessarily requires that the pressure in the interior 
of a star to be isotropic. This approach  has been used extensively in the 
study of polytropes, including white dwarfs, and of compact objects 
such as neutron stars \cite{GLENDENNING}.
However, theoretical advances in the last
decades indicate that, in many systems, deviations from local isotropy in the 
pressure, in particular at very high densities, may play an 
important role in determining their properties \cite{RUDERMAN,CANUTO}. 

The physical situations where anisotropic pressure may be relevant
are very diverse. By anisotropic pressure we mean that the radial component
of the pressure, $p_r(r)$, differs from the angular components, $p_{\theta}(r)
=p_{\varphi}(r)\equiv p_t(r)$. (That $p_{\theta}(r)=p_{\varphi}(r)$  is a
direct consequence of spherical symmetry.) Of
course, spherical symmetry demands both to be strictly a function of the radial
coordinate.
A scalar field with non-zero spatial gradient is an example of
a physical system where 
the pressure is clearly anisotropic. [This anisotropic character of a 
scalar field occurs  already at the level of special relativity, where it is
easy to show that $p_r-p_t = (d\phi/dr)^2$]. Boson stars, 
hypothetical self-gravitating compact objects resulting from the 
coupling of a complex scalar 
field to gravity, are systems where anisotropic pressure occurs naturally
\cite{B-STARS}. Similarly, the energy-momentum tensor of both electromagnetic
and fermionic fields are naturally anisotropic. Isotropy appears as an extra 
assumption on the behaviors of the fields or of the fluid modeling the stellar
interior.

In the interior of neutron stars pions may condense. 
It has been shown that due to the 
geometry of the ${\pi}^{-}$ modes, anisotropic distributions of pressure 
could be 
considered to describe a pion condensed phase configuration \cite{SAWYER}. The
existence of solid cores and type P superfluidity
may also lead to departures from
isotropy within the neutron star interior \cite{GLENDENNING}. Since
we still do not have a detailed microscopic formulation of the possible
anisotropic stresses emerging in these and other
contexts, we take the general approach 
of finding several exact solutions representing different physical
situations, modeled by {\it ansatze} for the anisotropy factor,
$p_t-p_r$. As a general rule, we find that the presence of anisotropy affects
the critical mass for stability, $2M/R$, and the surface redshift, $z_s$. These
physical consequences of pressure anisotropy are not new. Previous studies have
found some  exact solutions, assuming certain relations for the anisotropy 
factor \cite{BOWERS,PONCE,MEHRA,BONDI,CORCHERO,HERRER,HERRERA}.
Our goal here is to extend those results, offering a detailed analysis
of the changes in the physical properties of the stellar objects due to the
presence of anisotropy. Hopefully, our results will be of importance in the
analysis of data from compact objects, 
as well as in the study of the behavior of matter under strong gravitational 
fields.

This paper is organized as follows. In the next section, we set up
the equations used and the assumptions
made in our study. 
We restrict our exact solutions to two classes,
investigated in sections III and IV respectively. In section III, 
after reviewing the results of Bowers and Liang \cite{BOWERS}, we obtain
several new exact solutions for stars of constant density.
In section IV, we
examine solutions for the case $\rho(r) \propto 1/r^2$, which has been used to
model ultradense neutron star interiors \cite{WEINBERG}. 
We conclude in section V  with a
 brief summary of our results and an outlook to future work. 
In Appendix A, we demonstrate the equivalence of
the Tolman
and Schwarzschild masses for self-gravitating anisotropic spheres. In Appendix
B we present the general solution for stars featuring an energy density 
with a constant
part and a $r^{-2}$ contribution, as discussed in 
Section IV, in terms of hypergeometric functions.

\section{Relativistic Self-Gravitating Spheres}

\noindent We consider a static equilibrium distribution of matter 
which is spherically
symmetric. In Schwarzschild coordinates the metric can be written as 

\be
ds^2 = e^{\nu}dt^2  - e^{\lambda}dr^2 - r{^2}d{\theta}^2 - 
r{^2}\sin^{2}{\theta}d{\phi}^2~,
\ee
where all functions depend only on the radial coordinate $r$.
\noindent The most general energy-momentum tensor compatible with spherical
symmetry is
\be
T^{\mu}_{\nu} = {\rm diag}(\rho, -p_{r}, -p_{t}, -p_{t})~.
\ee
We see that isotropy is not required by spherical symmetry; it is an added
assumption.
\noindent The Einstein field equations for this spacetime geometry and matter 
distribution are
\be
\label{einstein.1}
e^{-\lambda} \left( \frac{{\nu}^{\prime}}{r}+ \frac{1}{r^2} \right)  - 
\frac{1}{r^2} = 8{\pi}p_{r}~;
\ee
\be
\label{einstein.2}
e^{-\lambda} \left( \frac{1}{2} {\nu}^{\prime \prime} - 
\frac{1}{4}{\lambda}^{\prime}{\nu}^{\prime} 
+ \frac{1}{4} \left({\nu}^{\prime} \right)^{2} +  \frac{\left({\nu}^{\prime} - 
{\lambda}^{\prime} \right)}{2r} \right) = 8{\pi}p_{t}~;
\ee

\be
\label{einstein.3}
e^{\lambda} \left( \frac{{\lambda}^{\prime}}{r} - \frac{1}{r^2} \right) + 
\frac{1}{r^2} = 8{\pi}{\rho}~.
\ee
\noindent Note that 
this is a system of 3 equations with 5 unknowns. Consequently, it is 
necessary to specify two equations of state, such as ${p_{r} = p_{r}(\rho)}$ 
and ${p_{t} = p_{t}(\rho)}$.

\noindent It is often
useful to transform the above equations into a form where the 
hydrodynamical properties of the system are more evident. 
For systems with isotropic 
pressure, this formulation results in the 
Tolman-Oppenheimer-Volkov (TOV) equation. 
The  generalized TOV equation, including
anisotropy, is 
\be
\frac{dp_{r}}{dr}  = -(\rho + p_{r})\frac{{\nu}^{\prime}}{2} + 
\frac{2}{r}(p_{t} - p_{r})~,
\ee
\noindent   with 
\be
\frac{1}{2}{\nu}^{\prime} = \frac{m(r) + 4 \pi r^{3} p_{r}}{r(r - 2m)},
\ee
\noindent and 
\be
m(r) = \int_{0}^{r} 4 \pi r^{2} \rho dr~.
\ee
\noindent Taking $r = R$ in the above expression gives us the
Schwarzschild mass, $M$. [This implicitly assumes that $\rho=0$ for $r>R$.]

\noindent A more general formula is the 
Tolman mass formula \cite{TOLMAN}:
\be
M = \int_{0}^{R}(2T^{0}_{0} - T_{\mu}^{\mu})(-g)^{\frac{1}{2}}r^{2}dr.
\ee
\noindent The equivalence of the Tolman and Schwarzschild masses 
for systems with anisotropic pressure is
demonstrated in Appendix A.
   
In order to solve the above equations we must impose appropriate boundary 
conditions. We require that the solution be regular at the origin. This 
imposes the condition that $m(r) \rightarrow 0$ as $r \rightarrow 0$. If 
$p_{r}$ is finite at the origin then ${\nu}^{\prime} \rightarrow 0$
as $ r \rightarrow 0$. The gradient $dp_{r}/dr$ will be finite at 
$r =0$ if $(p_{t} - p_{r})$ vanishes at least as 
rapidly as $r$ when $r \rightarrow 0$. This will be the case in all scenarios
examined here. 

The radius of the star is determined by the condition $p_{r}(R) =0$. It is 
not necessary for $p_{t}(R)$ to vanish at the surface. But it is reasonable to 
assume that all physically interesting solutions will have $p_{r} ,p_{t} \geq 
0 $ for $r \leq R$.  

\section{Exact Solutions For  $\rho = {\rm constant}$}

\noindent For $ \rho \equiv \rho_{0} = {\rm constant}$, we can write the 
generalized TOV equation as
\be
\frac{dp_{r}}{dr} =  \frac{2}{r}(p_{t} - p_{r}) - \frac{4 \pi r ( {p_{r}}^{2} +
 \frac{4}{3}p_{r}{\rho}_{0} + \frac{{{\rho}_{0}}^2}{3})}{ 1 - 
\frac{8}{3}\pi{\rho}_{0}r^2}~.
\ee    
\noindent Bowers and Liang \cite{BOWERS}
solved the generalized TOV equation by 
considering the following equation of state,
\be
p_{t} -p_{r} = C\frac{ r^2 ( {p_{r}}^{2} + 
\frac{4}{3}p_{r}{\rho}_{0} + 
\frac{{{\rho}_{0}}^2}{3})}{ 1 - \frac{8}{3}\pi{\rho}_{0}r^2}~,
\ee
\noindent that is, they considered the term related to the anisotropy to be 
simply proportional to the usual hydrodynamic term on the right hand side. The
constant parameter $C$, which we will also use, measures the amount of 
anisotropy.

They found that the radial pressure is given by 
\be
p_{r} = {\rho}_{0} \left[ \frac{(1 - 2m/r)^{Q} - 
(1 - 2M/R)^{Q}}{3(1 - 2M/R)^{Q} -(1 - 2m/r)^{Q}}  \right]~,
\ee
\noindent where the total mass $M \equiv m(R)$, $R$ is the radius of the system,
and $Q = \frac{1}{2}  - \frac{3C}{4\pi}$. The central pressure is given by
\be
\label{central_pressure}
p_{c} = {\rho}_{0}\left [ \frac{1 - (1 - 2M/R)^{Q}}{3(1 - 2M/R)^{Q} - 1}
\right ]~.
\ee
\noindent An equilibrium configuration exists for 
all values of ${2M}/R$ such that  
$p_{c}$ is finite. The critical model results for that value of ${2M}/R$
such that $p_{c}$ becomes infinite. From eq. \ref{central_pressure},
this occurs when the denominator vanishes, that is, when
\be
(2M/R)_{ \rm {crit}} = 1 - (\frac{1}{3})^{\frac{2}{1 - \xi}}~,
\ee
\noindent where $\xi\equiv 3C/2$.
We see that for $ 0 \leq \xi < 1,$ 
$(2M/R)_{{\rm crit}}$ can be greater than $8/9$,
the maximum value for an isotropic configuration \cite{WEINBERG}.
\noindent For a given $\rho_0$ and $C$, the critical mass is 
\be
\label{critical_mass1}
M_{{\rm crit}} = (\frac{3}{32 \pi \rho_0})^{\frac{1}{2}}
\left [ 1 - (\frac{1}{3})^{\frac{2}{1 - \xi}}\right ]^{\frac{3}{2}}~.
\ee
\noindent Equation \ref{critical_mass1}
shows that the critical mass is less 
than the isotropic value when $C < 0$. 
When $C > 0$ the critical mass exceeds the isotropic limit. 
For a given ${\rho}_{0}$, the maximum value of the 
ratio of the critical mass to the isotropic mass approaches
\be
M_{a}(\xi =1)/M_{i} \simeq 1.19~,
\ee
where $M_{a}(\xi = 1)$ corresponds to a configuration uniformly filling up 
to its own Schwarzschild radius. This represents a maximum of 19\% 
increase in the stable mass, comparable with results for relativistic 
models 
of slowly rotating isotropic stars
\cite{THORNE}.
\noindent The increase in stellar mass also affects the surface redshift, 
\be
z_s = (1 - {2M}/R)^{-\frac{1}{2}} - 1~.
\ee
\noindent For the Bowers-Liang solution, the maximum redshift is,
\be
z_{{\rm c}} = 3^{\frac{1}{1 - \xi}} - 1.
\ee
\noindent For $\xi = 0$ 
we recover the well-known isotropic result $z_{\rm {c}} = 2$. The 
introduction of anisotropy removes the upper limit, since as $\xi \rightarrow
1$, 
$z_{{\rm c}}$ can be
arbitrarily large. Thus, in principle, a modest amount of 
anisotropy is capable of 
explaining surface redshifts greater than 2, an intriguing possibility. 

For stars with constant energy density $\rho_{0}$, we 
considered solutions to two general ${\it ansatze}$, which we label cases I and
II for convenience. We begin with case I:

\noindent {\bf CASE I}: The anisotropy factor is written as
\be
p_{t} - p_{r}  = C r^2F(p_{r},{\rho}_{0})
\left(1 - \frac{8}{3}\pi {\rho}_{0} r^{2} \right)^{-1}~, 
\ee
where $C$ measures the amount of anisotropy and 
the function $F(p_{r},{\rho}_{0})$ includes 6 separate cases,
\be
F = ( {p_{r}}^{2};p_{r}{\rho}_{0};{{\rho}_{0}}^{2} ;
{p_{r}}{^2}  + p_{r}{\rho}_{0};{p_{r}}{^2}  + {{\rho}_{0}}^{2} ;
p_{r}{\rho}_{0} + {{\rho}_{0}}^{2})~.       
\ee

For the sake of brevity, we will only discuss 2 of the 6 possible cases. 
It should be straightforward to obtain solutions for the other cases.

{\bf Case I.1}  $F(p_{r},{\rho}_{0})={\rho_{0}}^2$ \\

When  $F(p_{r},{\rho}_{0})={\rho_{0}}^2$ the $TOV$  equation becomes
\be
\frac{dp_{r}}{dr} =   - \frac{4 \pi r
 \left[ {p_{r}}^{2} +
\frac{4}{3}p_{r}{\rho}_{0} + ( \frac{1}{3} - \frac{C}{2 \pi}){{\rho}_{0}}^2
\right]}
{ 1 - \frac{8}{3}\pi{\rho}_{0}r^2}.
\ee    
Thus we find
\be
\int_{0}^{p_{r}} \frac{d p_{r}}{  \left[ {p_{r}}^{2} +
\frac{4}{3}p_{r}{\rho}_{0} + ( \frac{1}{3} - \frac{C}{2 \pi}){{\rho}_{0}}^2
\right]}
 = - 4 \pi \int_{R}^{r} \frac{dr}{ 1 - \frac{8}{3}\pi{\rho}_{0}r^2}.
\ee
The solutions naturally divide into 3 subcases, according to the value of
the anisotropy parameter, which we write as $ k \equiv{C}/({2 \pi})$. 

When $k > -1/9$ the radial pressure is given by 
\be
p_{r}(r) = {{\rho_0}} \left[ \frac{(1 - 3k)[(1 - 2m/r)^{\frac{A}{2}} - 
(1 - 2M/R)^{\frac{A}{2}}]}{( 2 + A) 
(1 - 2M/R)^{\frac{A}{2}} - 
( 2 - A)(1 - 2m/r)^{\frac{A}{2}}}  \right]~,
\ee
\noindent where $ A \equiv ( 1 + 9k)^{\frac{1}{2}}$. Clearly, when  $k=0$ we
recover the isotropic limit. In figure 1.1, we show
the radial pressure as a function of the radius parameterized in terms of
fractions of the critical configuration, $(2M/R)_{{\rm crit}}$.
The solutions are normalized to $p_{r}(R)=0$. 
Alternatively, to obtain a solution we can fix $\rho_0$ and $p_c$, the
central pressure, and compute the value of $r$ for which $p_r(R)=0$.

\hspace{2.0in}
\psfig{figure=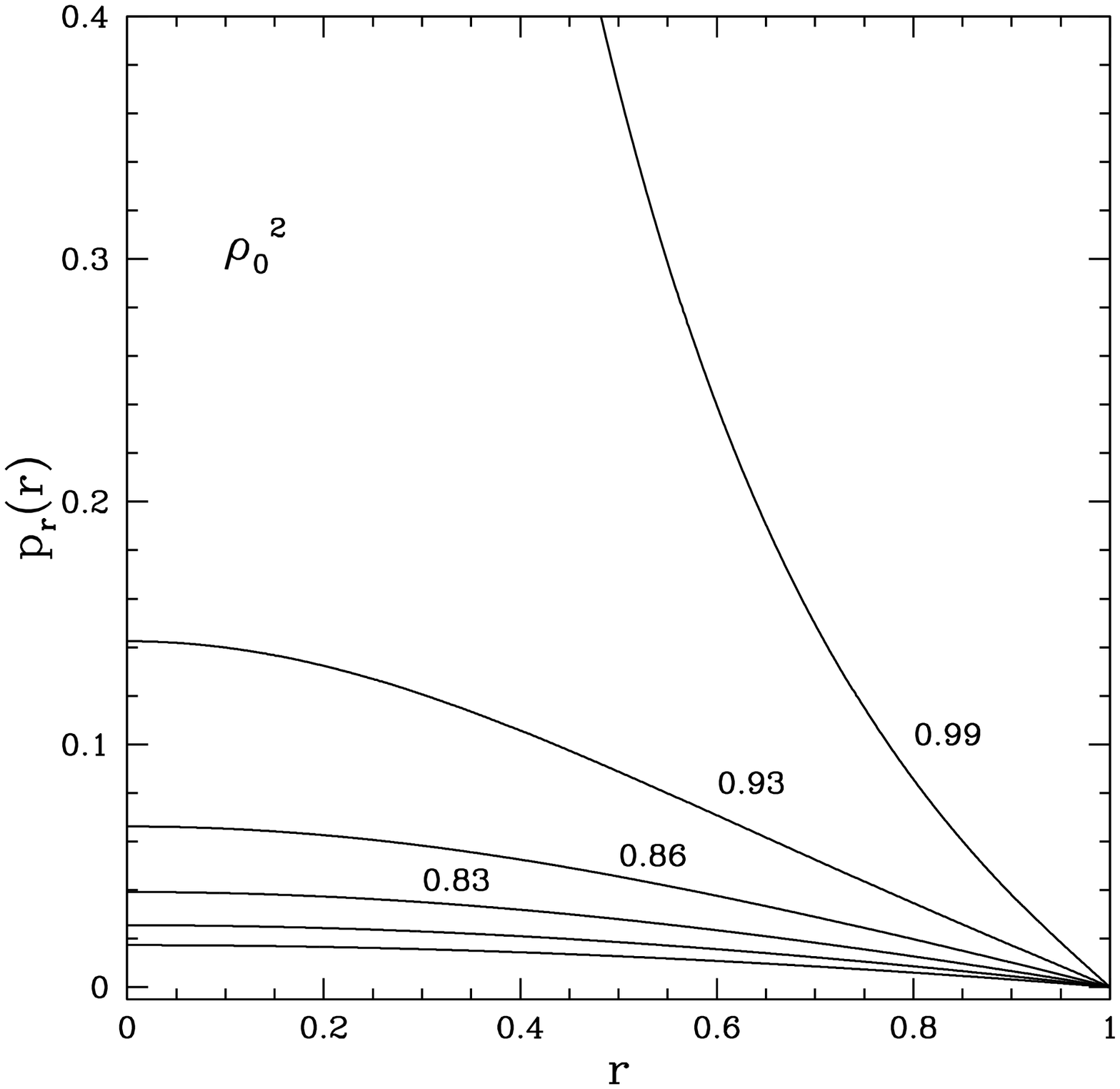,width=2.7in,height=2.5in}

\setlength{\baselineskip}{5mm}

\noindent Fig. 1.1: Radial pressure as a function of radius 
for the case $ k = 1/2\pi$, $ \rho = const $ and $ (p_{t} - p_{r}) 
\propto \rho^{2}$ parameterized in terms 
of fractions of the critical configuration $(2M/R)_{crit}$.

\setlength{\baselineskip}{8.25mm}

The central pressure is given by
\be
p_{c} = {{\rho_0}} \left [ \frac{( 1 - 3k)[ 1 -
(1 - 2M/R)^{\frac{A}{2}}]}{( 2 + A) 
(1 - 2M/R)^{\frac{A}{2}} - ( 2 - A) } \right ]~.
\ee
\noindent Thus, for $ k > -1/9$ we obtain,
\be
\left ( \frac{2M}{R} \right  )_{\rm{crit}} = 1 
 - \left [ \frac{(2 -A)}{(2 + A)}\right ]^{\frac{2}{A}}~.
\ee
\noindent For $ |k| \ll 1$ we find that, correcting slightly Bowers
and Liang \cite{BOWERS},
\be
\left ( \frac{2M}{R} \right )_{\rm{crit}}
 \simeq \frac{8}{9} + \left (\frac{4}{3} - \ln 3\right )k 
+{\cal O}(k^2)~.
\ee 
\noindent We see that a positive anisotropy, $k>0$, leads to a violation
of the isotropic limit $({2M}/{R})_{{\rm crit}} = {8}/{9}$. 
However, as $k\rightarrow 1/3$, $(2M/R)_{{\rm crit}}
\rightarrow 1$ and $p_r < 0$: 
there is a maximum allowed anisotropy in this case.
When $ k < 0 $,  $ ({2M}/{R})_{ \rm {crit}} $ is always less than $ {8}/{9}$.

The maximum surface redshift for $k > -1/9$ is 
\be
z_{{\rm {c}}} =  \left (\frac{2 + A}{2 -A}\right )^{\frac{1}{A}} - 1~.
\ee

For $ k = {-1}/{9}$, the radial pressure is 
\be
p_{r}(r) = \frac{2}{3}{\rho}_{0} \left [ \frac{1}{1 - 
{1\over 2}\ln(\frac{1 - 2m/r}{1 - 2M/R})} -1 \right ]~.
\ee
\noindent This matches smoothly the result for $k > -1/9$, as $k \rightarrow
-1/9$.

When $ k < -1/9$ the radial pressure is given by
\be
p_{r}(r) = {{\rho_0}\over 3}\left \{B\tan\left [ \arctan \left (\frac{2}{B}
\right ) + 
\frac{B}{4} \left (\ln\left ({{1 - 2m/r}\over {1 - 2M/R}}\right )\right )
 \right ] - 2\right \}~,
\ee 

\noindent with $B \equiv (9|k| - 1)^{\frac{1}{2}}$. Now, the central pressure 
is
\be
p_{c} ={{\rho_0}\over 3}\left [B\tan\left (\arctan\left(\frac{2}{B}\right) 
- \frac{B}{4} 
\ln(1 - 2M/R)\right )  - 2 \right ]~,
\ee
\noindent and the critical values of $2M/R$ are
\be
\left (\frac{2M}{R}\right )_{\rm {crit}}
 = 1 - \exp\left [ \frac{4}{B}\arctan\left(\frac{2}{B}\right) - 
\frac{\pi}{2} \right ]~.
\ee
In figure 1.2, we plot the critical values of $2M/R$ as a function 
of the anisotropic parameter $C$. 

The maximum surface redshift for $ k < {-1}/{9}$  is 
\be
z_{{\rm {c}}} = \exp \left [\frac{\pi}{4} 
 -  \frac{2}{B}\arctan(\frac{2}{B}) \right ] -1~.
\ee

In figure 1.3, we plot the maximum surface redshift as a function of 
the anisotropic parameter $C$. We find that for $ A \rightarrow 2$, 
{\it i.e.}, $ k \rightarrow  1/3$, the surface redshift becomes infinite. 
Thus, in this model, positive anisotropies generate arbitrarily 
large surface redshifts.

\hspace{2.0in}
\psfig{figure=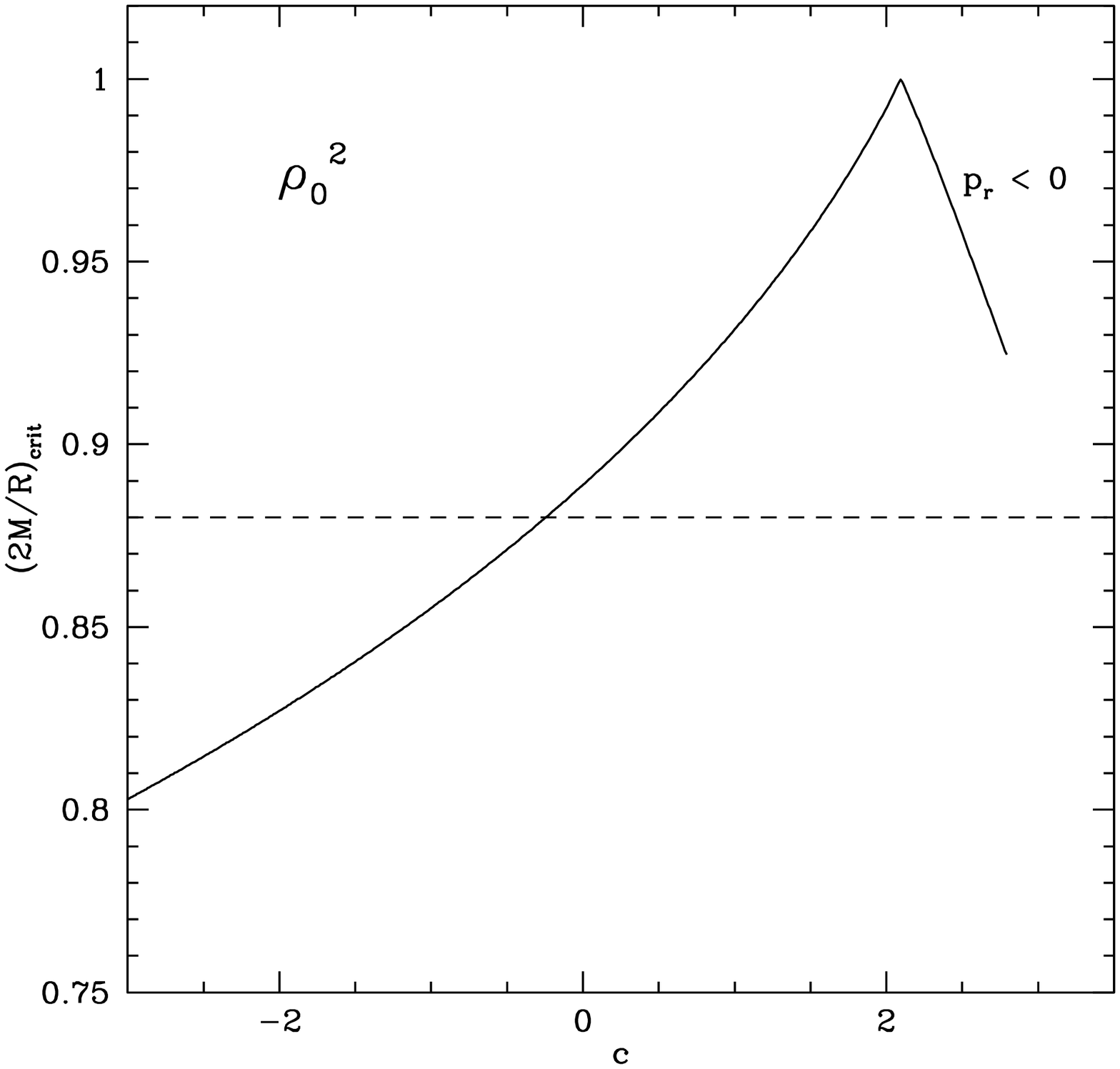,width=2.7in,height=2.5in}

\setlength{\baselineskip}{5mm}

Fig. 1.2  Critical values of $2M/R$ as a function of the 
anisotropic parameter $C$
 for  the case $ \rho = const $ and $ (p_{t} - p_{r}) \propto \rho^{2}$.

\setlength{\baselineskip}{8.25mm}

\hspace{2.0in}
\psfig{figure=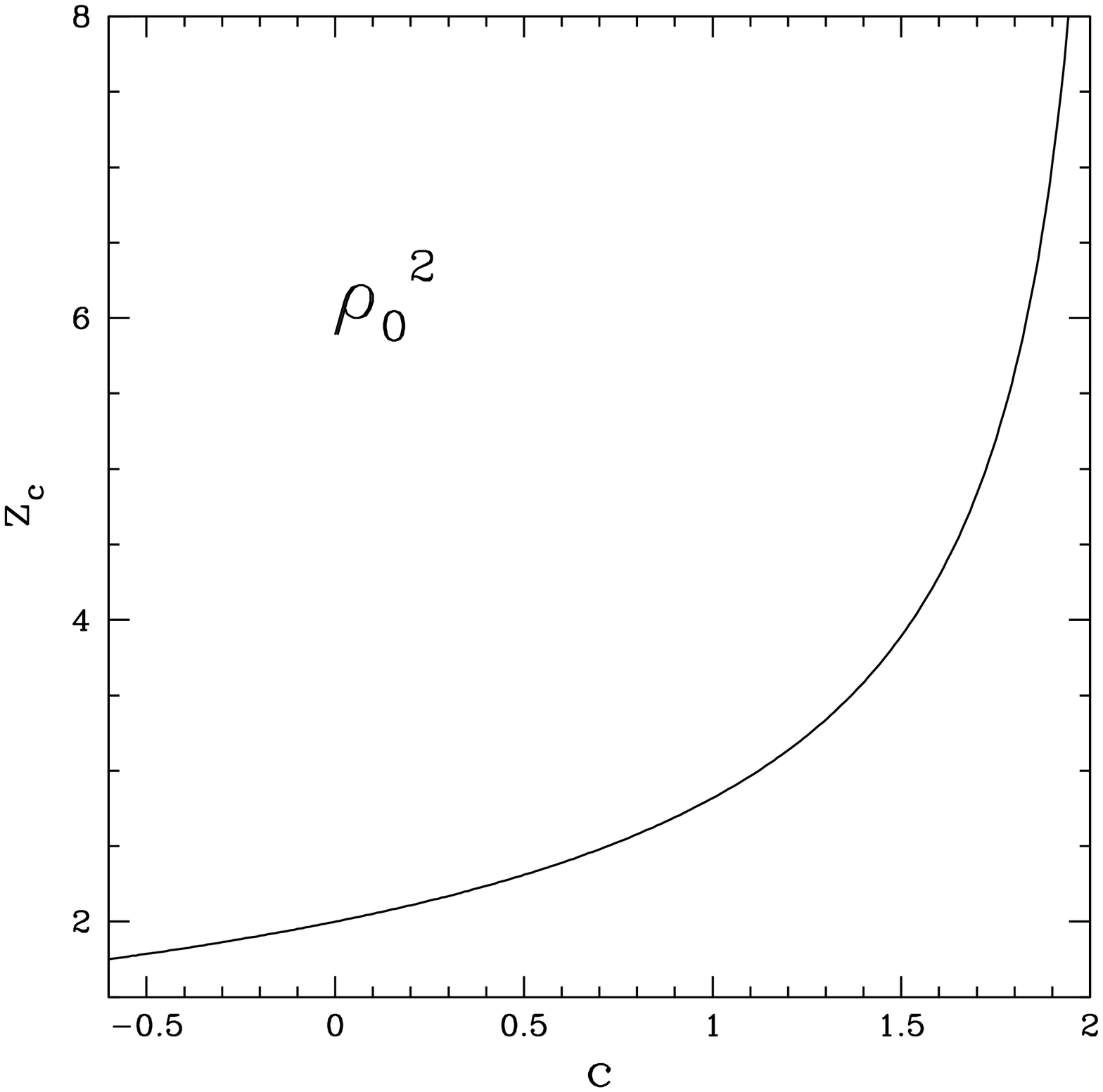,width=2.7in,height=2.5in}

\setlength{\baselineskip}{5mm}
Fig 1.3: Maximum surface  redshift as a function of the 
anisotropic parameter $C$
 for  the case $ \rho = const $ and $ (p_{t} - p_{r}) \propto \rho^{2}$.

\setlength{\baselineskip}{8.25mm}

\noindent {\it {\bf Case I.2}}: $F(p_r,\rho_0)=p_r^2$ 

This solution also separates into 3 subcases, 
depending on the value of the anisotropy parameter, $k=C/(2\pi)$.
When $ k > - 1/3$, the radial pressure is given by
\be 
p_{r}(r) = {\rho}_{0} \left [ \frac{(1 - 3k)[(1 - 2m/r)^{\frac{A}{2}} - 
(1 - 2M/R)^{\frac{A}{2}}]}{( 2 +  A) 
(1 - 2M/R)^{\frac{A}{2}} - 
( 2 - A)(1 - 2m/r)^{\frac{A}{2}}}  \right ]~,
\ee
where $ A \equiv ( 1 + 3k)^{\frac{1}{2}}$. In figure 2.1, we show
the radial pressure as a function of the radial coordinate for fractions 
of $(2M/R)_{{\rm crit}}$ and $ k = 1/\pi$.

\hspace{2.0in}
\psfig{figure=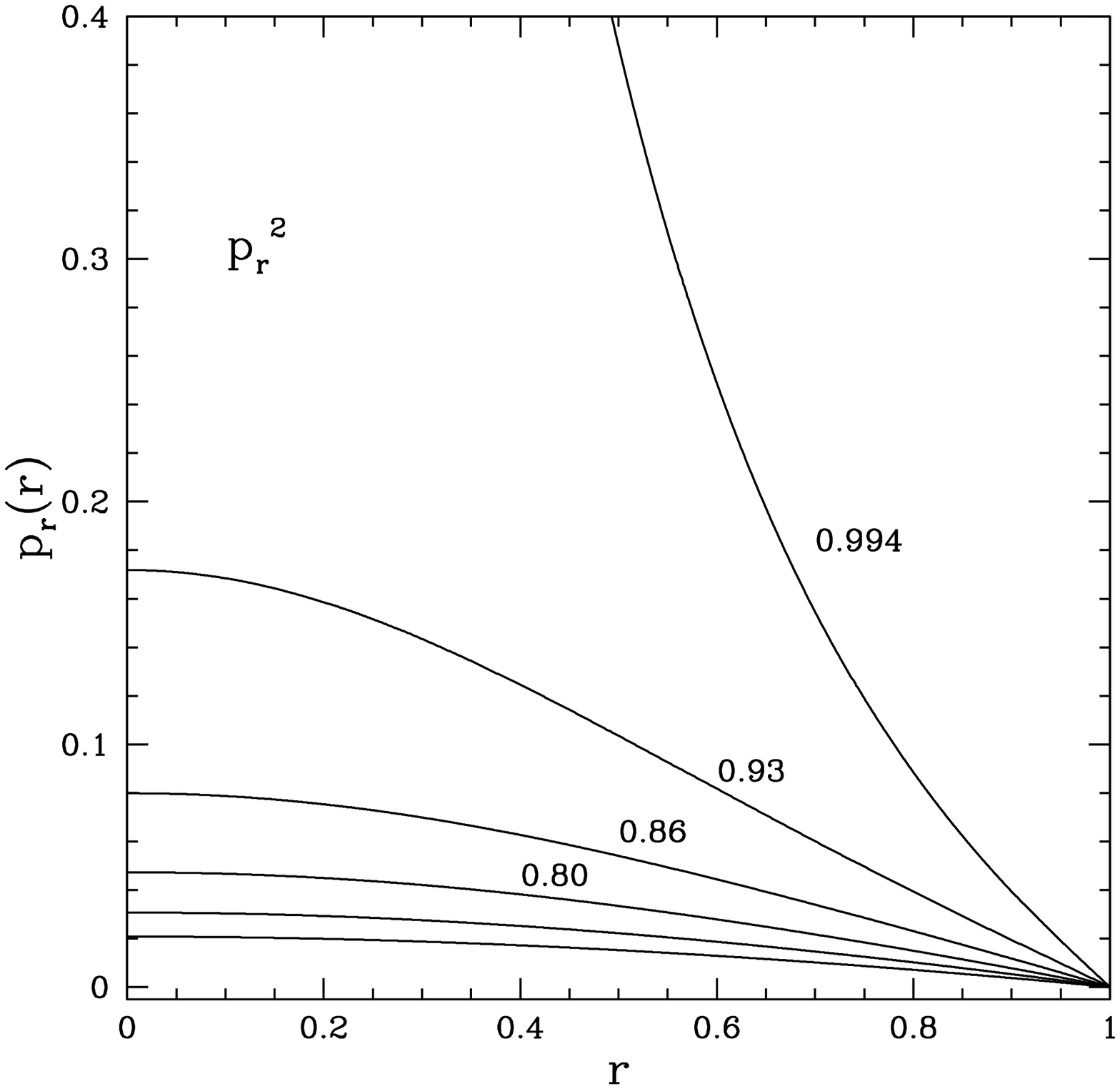,width=2.7in,height=2.5in}

\setlength{\baselineskip}{5mm}
\noindent Fig 2.1:  Radial pressure as a function of radius for the case 
 $ k = 1/\pi$, $ \rho = const $ and $ (p_{t} - p_{r}) \propto p_{r}^{2}$ 
parameterized in terms 
of fractions of the critical configuration $(2M/R)_{crit}$.

\setlength{\baselineskip}{8.25mm}

The central pressure is given by 
\be
p_c = {\rho}_{0} \left [ \frac{ (1 -3k)[1- 
(1 - 2M/R)^{\frac{A}{2}}]}{( 2 + A) 
(1 - 2M/R)^{\frac{A}{2}} - 
( 2 - A)}  \right ]~.
\ee
The critical configurations for the anisotropy parameter $k > -1/3$
 are given by 
\be
\left ( \frac{2M}{R}\right )_{\rm {crit}} = 1 
- \left [ \frac{(2 -A)}{(2 + A)}\right ]^{\frac{2}{A}}~,
\ee 
and the corresponding  maximum redshift for these values of $k$ are
\be
z_{{\rm {c}}} = 
\left [ \frac{(2 +A)}{(2 - A)}\right ]^{\frac{1}{A}} - 1~.
\ee

For $ k = -1/3$,
the radial pressure is given by
\be
p_{r}(r) = {\rho}_{0} \left [ \frac{1}{2 - 
\ln(\frac{1 - 2m/r}{1 - 2M/R})} - \frac{1}{2} \right ]~,
\ee
and the central pressure is 
\be
p_c ={\rho}_{0} \left [ \frac{1}{2 + \ln(1 - 2M/R)} -\frac{1}{2}  \right ]~.
\ee 
\noindent Thus, for $k = -1/3$ we have
\be
\left ( \frac{2M}{R} \right )_{\rm{crit}} = 1 - e^{-2}~.
\ee

The solution for $ k < -1/3$ is 
\be 
p_{r}(r) = {{\rho_0}\over 3( 1 - k)}\left \{B\tan\left [ \arctan \left (
\frac{2}{B}
\right ) + 
\frac{B}{4} \ln\left ({{1 - 2m/r}\over {1 - 2M/R}}\right )
 \right ] - 2\right \}~,
\ee 
with $ B \equiv (3|k| -1)^{\frac{1}{2}}$. The central pressure is 
\be
p_{c} = \frac{{\rho}_{0}}{3(1-k)}
\left [B\tan\left (\arctan\left(\frac{2}{B}\right) - \frac{B}{4} 
\ln(1 - 2M/R)\right )  - 2 \right ]~.
\ee
Now,
\be
\left ( \frac{2M}{R} \right )_{\rm{crit}} = 
1 - \exp\left [ \frac{4}{B}\arctan(\frac{2}{B}) - 
\frac{\pi}{2} \right ]~,
\ee
\noindent and the corresponding maximum surface redshift is
\be
z_{\rm{c}} = \exp \left[ \frac{\pi}{4} 
 -\frac{2}{B}\arctan\left(\frac{2}{B}\right) \right] - 1 ~.
\ee
Note that the critical values of $2M/R$ and surface redshifts for cases I.1
and I.2 are identical, up to a shift in $k \rightarrow k/3$. 
In figures 2.2  and 2.3 we plot the critical values of $2M/R$ 
and the  maximum surface redshift as a function of 
the anisotropic parameter $C$.

\hspace{2.0in}
\psfig{figure=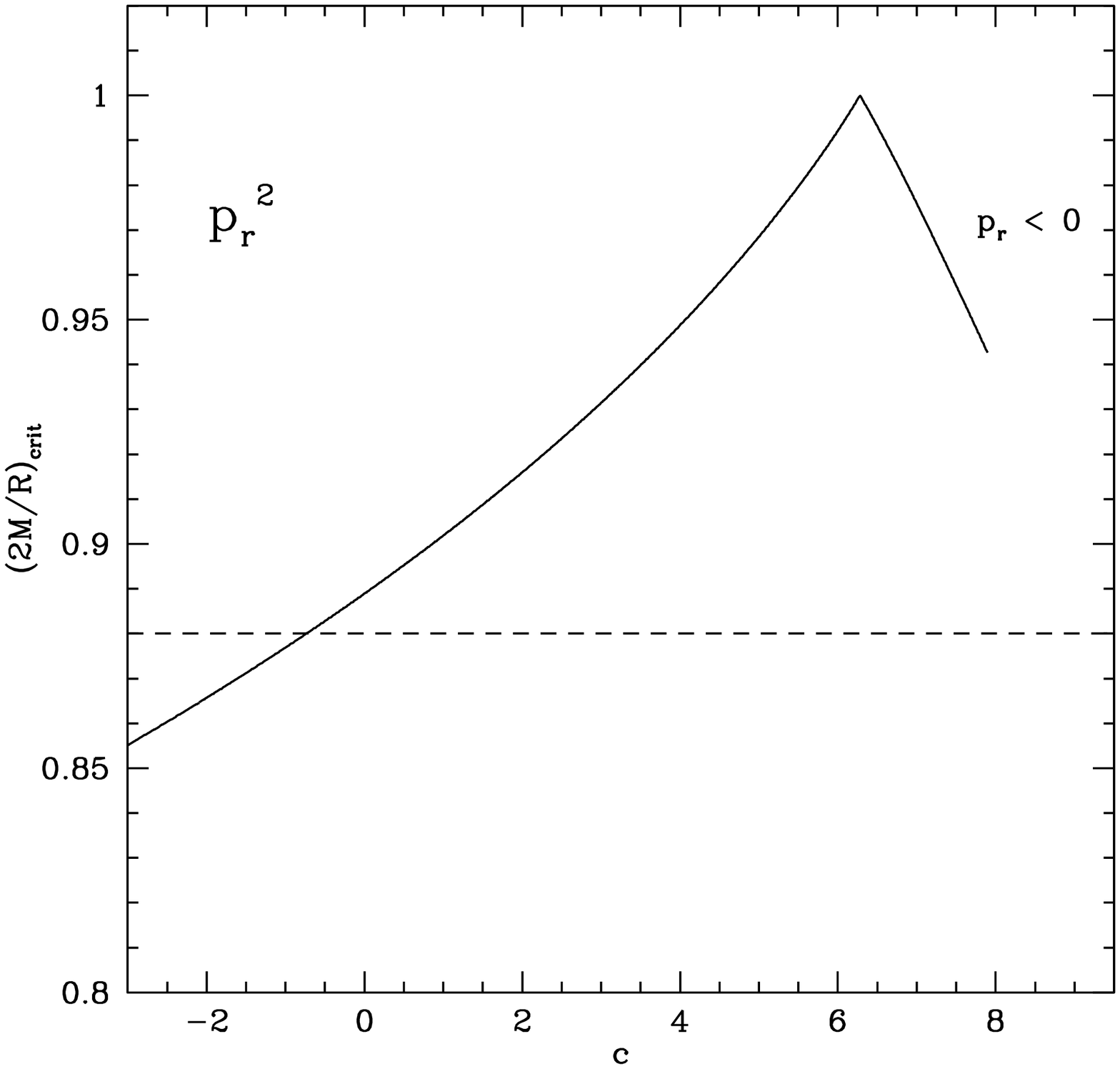,width=2.7in,height=2.5in}

\setlength{\baselineskip}{5mm}
\noindent Fig 2.2: Critical values of $2M/R$  as a function of 
the anisotropic parameter $C$
 for  the case $ \rho = const $ and $ (p_{t} - p_{r}) \propto p_{r}^{2}$.

\setlength{\baselineskip}{8.25mm}

\hspace{2.3in}
\psfig{figure=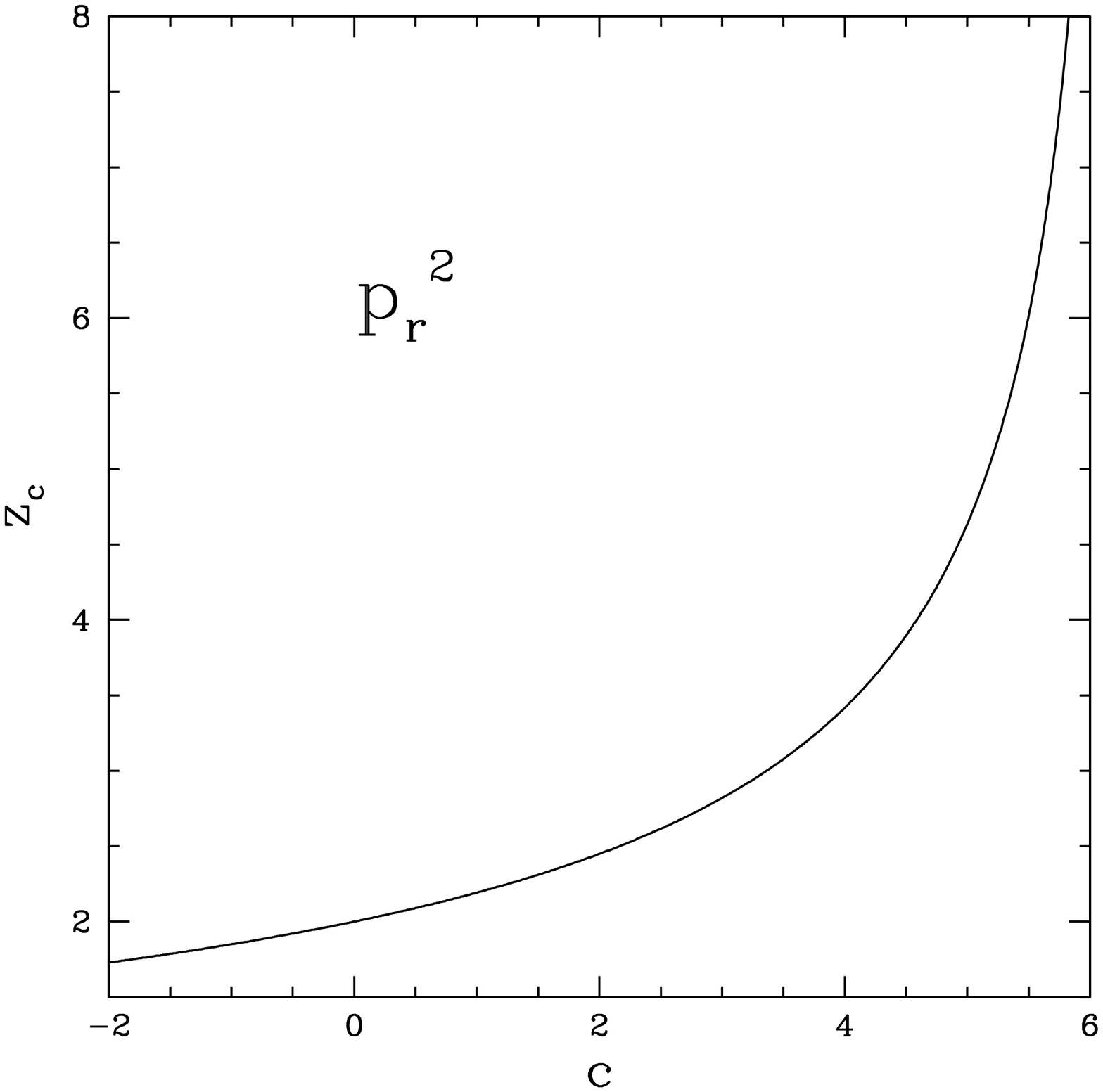,width=2.7in,height=2.5in}

\setlength{\baselineskip}{5mm}
\noindent Fig 2.3:  Maximum  surface redshift  as a 
function of the anisotropic parameter $C$
 for  the case $ \rho = const $ and $ (p_{t} - p_{r}) \propto p_{r}^{2}$.

\setlength{\baselineskip}{8.25mm}

\vspace{1.cm}

\noindent{\bf CASE II}

\noindent The second class of exact solutions with constant density
 follows the 
{\it ansatze},
\be
\label{exponential}
p_{t} - p_{r}  = \frac{C}{\rho_0}\frac{{\frac{r^n}{R^n}} 
\exp(-\frac{r}{R}) ( {p_{r}}^{2} + 
\frac{4}{3}p_{r}{\rho}_{0} + 
 \frac{{{\rho}_{0}}^2}{3})}{ 1 - \frac{8}{3}\pi{\rho}_{0}r^2}~,
\ee
and
\be
\label{Gaussian}
p_{t} - p_{r}  = \frac{C}{\rho_0}\frac{{\frac{r^n}{R^n}} 
\exp(-\frac{r^2}{R^2}) ( {p_{r}}^{2} + 
\frac{4}{3}p_{r}{\rho}_{0} + 
\frac{{{\rho}_{0}}^2}{3})}{ 1 - \frac{8}{3}\pi{\rho}_{0}r^2}~,
\ee
where $n\ge 2$ is an integer.
The motivation for this choice of anisotropy comes from 
 boson stars \cite{B-STARS}, where  it is found that the anisotropy 
factor vanishes at the origin and outside the star, reaching a maximum 
somewhere around the approximate radius of the configuration. (Boson stars
do not have a sharp boundary between the inside and the outside, as the scalar
field vanishes exponentially for $r > R$. One may think of it as a diffuse
``atmosphere'' around the denser stellar core.) 

For all cases considered, we found that there are values of 
$C>0$ for which $(2M/R)_{{\rm crit}}$ can be greater 
than $8/9$ and $z_{{\rm c}}$ can
approach arbitrarily large values. 
Let us focus on
the case where the anisotropy falls exponentially with distance, 
as in eq. \ref{exponential}. The case for the ``Gaussian'' anisotropy 
(eq. \ref{Gaussian}) can be solved
by following the same procedure. After integration we obtain,
\be
p_r(r) = \rho_0\frac{Z(r) - 1}{3 - Z(r)}~,
\ee
where
\be
Z(r) \equiv \left (\frac{1-8\pi\rho_0r^2/3}{1-8\pi\rho_0R^2/3}\right )^
{1\over 2} \exp\left [\frac{4C}{3R^n}\int_R^r r'^{(n-1)
}\exp(-r'/R)dr' \right ]~. 
\ee
 For a given $n$ the integral can be easily performed and we obtain an
expression for the radial pressure. Note that for physically acceptable 
solutions (${\it i.e.}$, with $p_r(r)>0$)
the function $Z(r)$ must satisfy $1\le Z(r) \le 3$. In figure 3.1 we show
the radial pressure as a function of radial distance for various fractions
of $(2M/R)_{{\rm crit}}$ and 
$n=2$.  In figure 3.2, we plot the anisotropy factor for the same fractions of
$(2M/R)_{{\rm crit}}$ and $n=2$.

\hspace{2.0in}
\psfig{figure=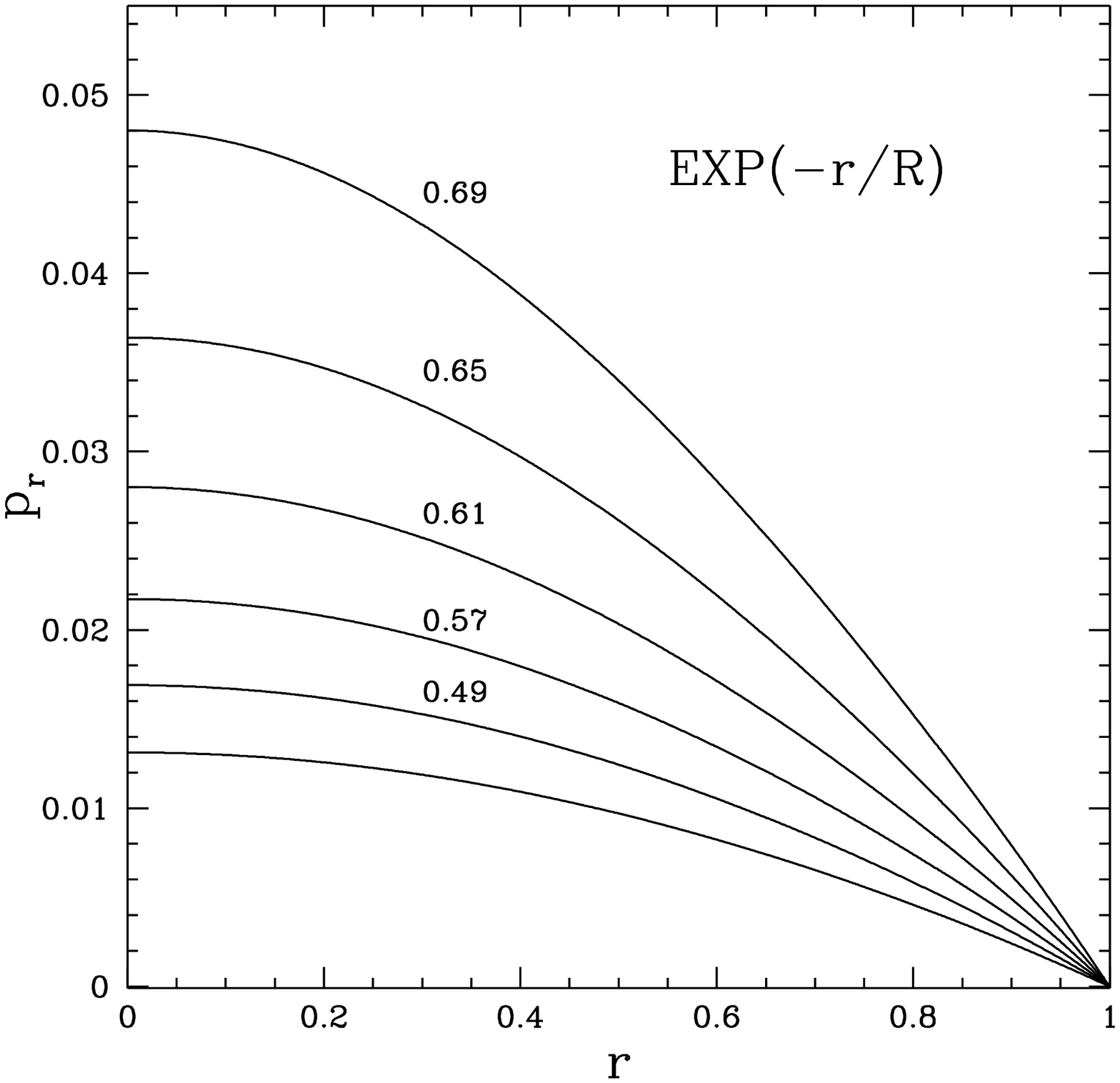,width=2.7in,height=2.5in}

\setlength{\baselineskip}{5mm}

\noindent Fig. 3.1: Radial pressure as a function of radius for the case 
$ \rho = const $ and $ p_{t} - p_{r} \propto 
(\frac{r}{R})^{2} \exp(-\frac{r}{R})$  parameterized in terms 
of fractions of the critical configuration $(2M/R)_{crit}$.

\setlength{\baselineskip}{8.25mm}

\hspace{2.0in}
\psfig{figure=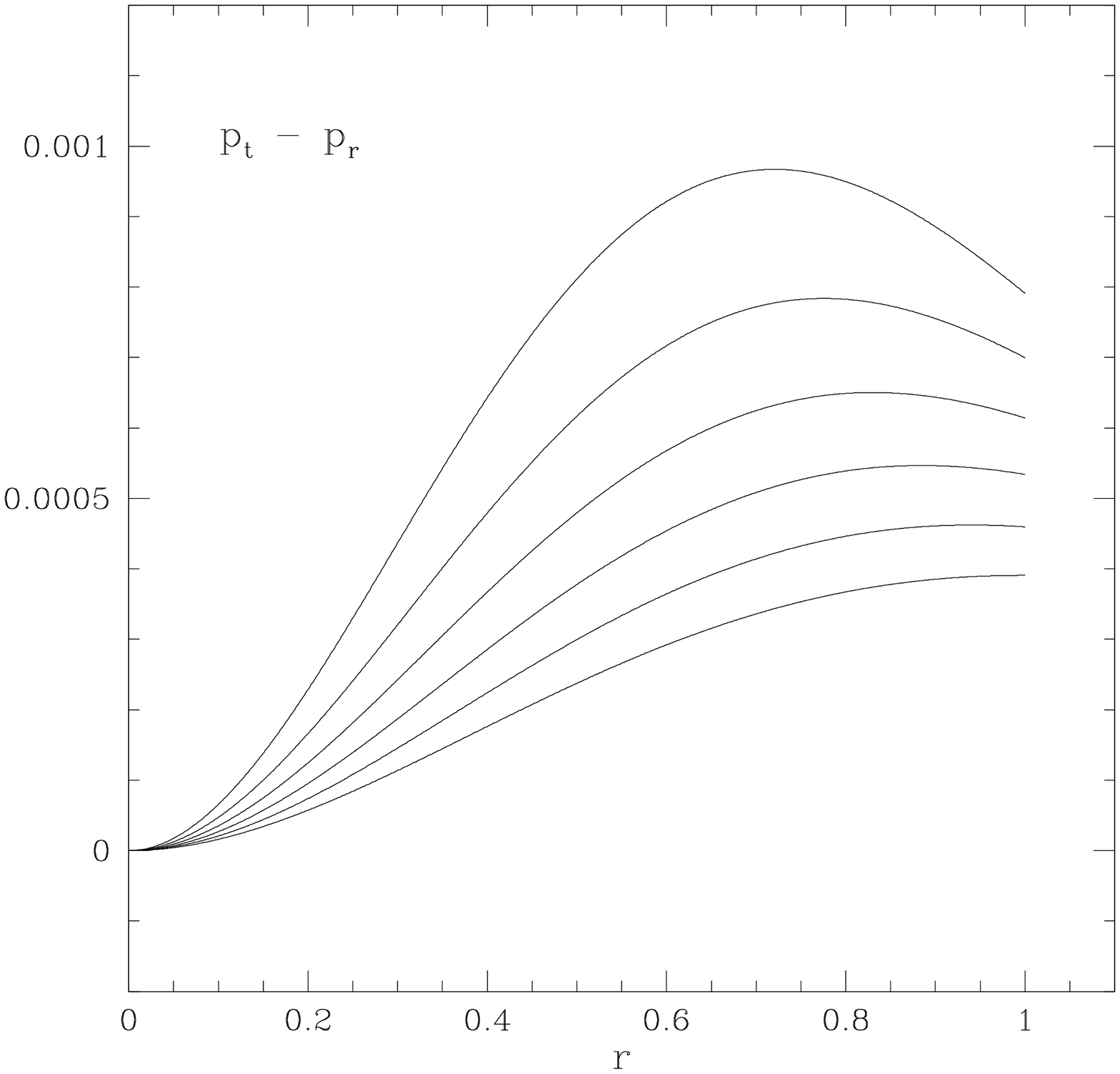,width=2.7in,height=2.5in}

\setlength{\baselineskip}{5mm}
\noindent Fig. 3.2: Anisotropy  factor as a function of the radius 
for the  same case  as in fig. 3.1.

\setlength{\baselineskip}{8.25mm}

Of more interest is the behavior at the
origin. From the above solution, the central pressure can be written as,
\be
p_c = \rho_0\frac{Z(0) - 1}{3 - Z(0)}~,
\ee
where
\be
Z(0) = \left (1-8\pi\rho_0R^2\right )^{-1/2}
\exp\left [\frac{4C}{3R^n}\int_R^0 r'^{(n-1)}\exp(-r'/R)dr' \right ]~.
\ee
The critical configuration is obtained for $Z(0) =3$. For example, for $n=2$
we obtain,
\be
\left ( \frac{2M}{R}\right )_{\rm{crit}} =
1 - {1\over 9}\exp\left [-{{8C}\over 3}\left (1-2e^{-1} \right )\right ]~.
\ee
Note that, as $C\rightarrow \infty$, $(2M/R)_{crit} \rightarrow 1$: there is no
maximum positive anisotropy. On the other hand, there is
a maximum negative anisotropy, beyond which the central pressure becomes 
negative 
($Z(0) < 1$).
This is given by $|C| = {3\over {8(1-2e^{-1})}}\ln 9 \simeq 3.12$. In
the Introduction, we noted
that boson stars have negative pressure anisotropy, $p_t - p_r < 0$. 
It is an interesting  open question if such 
``exploding'' solutions with  negative core pressures represent a new kind of 
instability of bosonic stellar configurations. 

In figure 3.3 we plot the critical mass as a 
function of anisotropy, for several values of $n$. 
In figure 3.4 we do the same for the surface redshift. For $n =2$ the surface 
redshift is given by, 
\be z_{\rm{c}} = 
3\exp\left [ {4C\over 3}\left (1-2e^{-1}  \right )\right ] -1~.  
\ee

\hspace{2.0in}
\psfig{figure=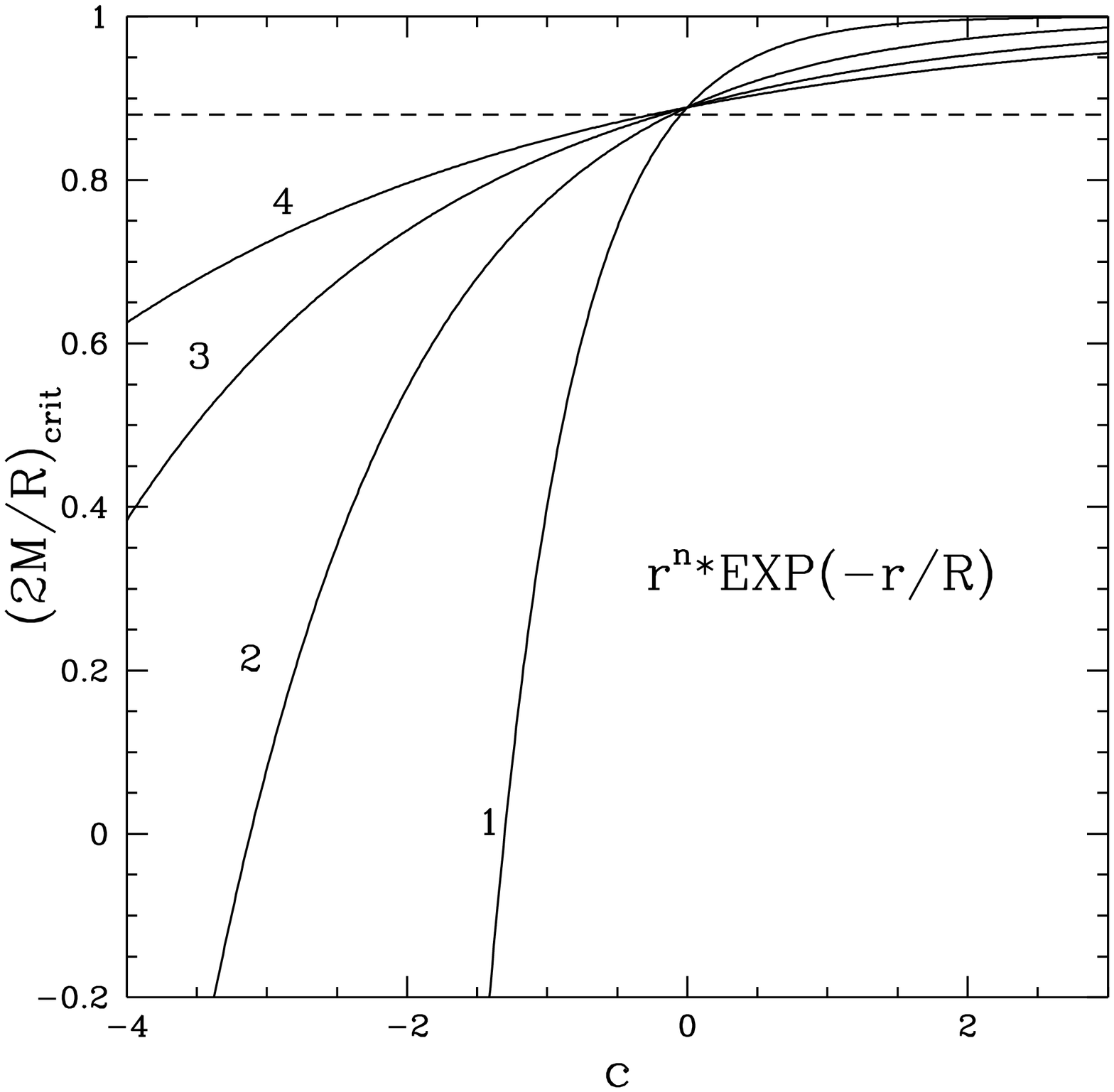,width=2.7in,height=2.5in}

\setlength{\baselineskip}{5mm}
\noindent Fig. 3.3:  Critical values of $2M/R$ as a function of 
anisotropy for the case 
$ \rho = const $ and $ p_{t} - p_{r} \propto 
(\frac{r}{R})^{n} \exp{-(\frac{r}{R})}$ with $ n = 1,2,3,4$.

\setlength{\baselineskip}{8.25mm}

\hspace{2.0in}
\psfig{figure=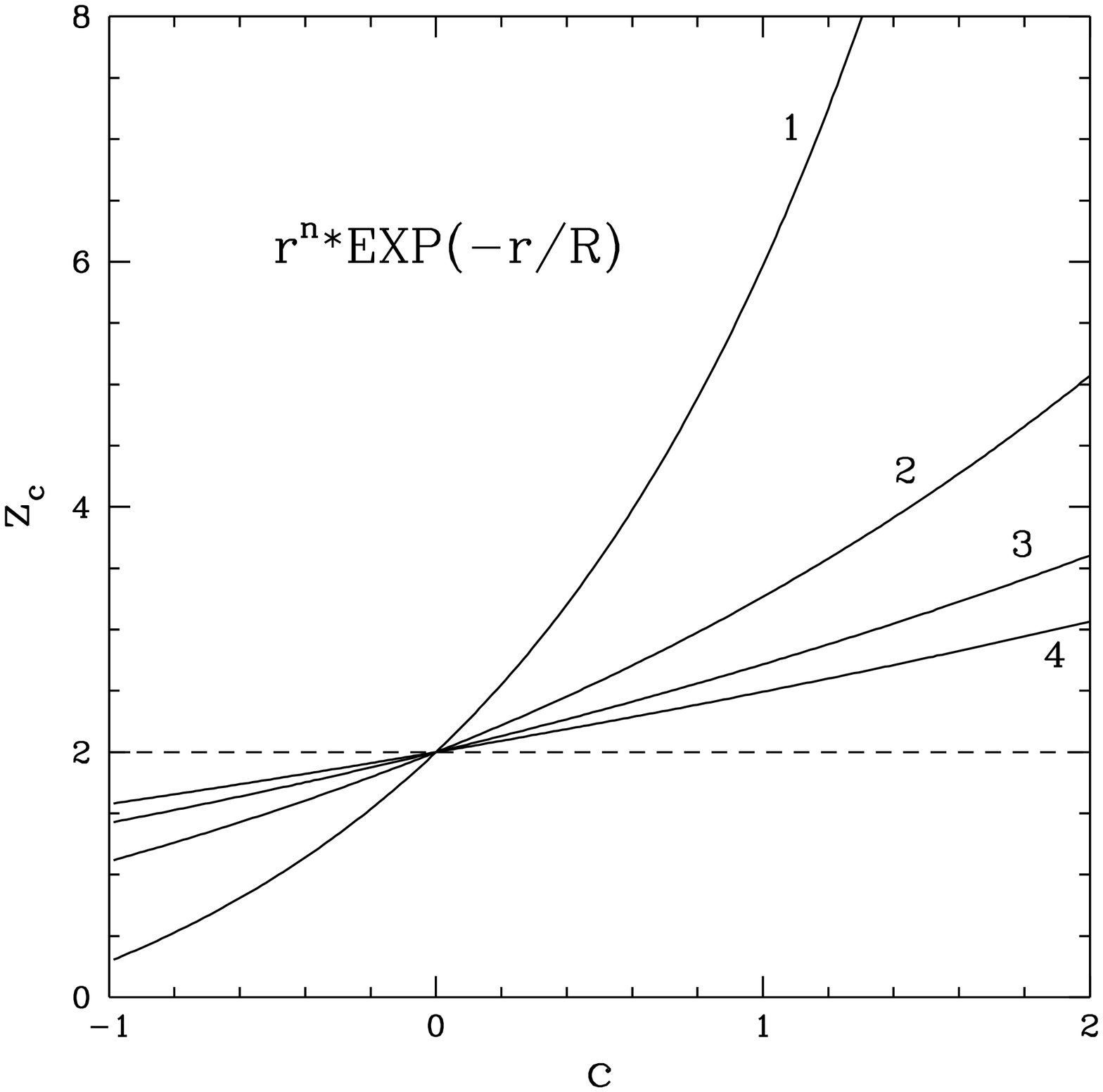,width=2.7in,height=2.5in}

\setlength{\baselineskip}{5mm}
\noindent Fig. 3.4:  Maximum surface redshift  as a function of anisotropy 
for the case $ \rho = const $ and $ p_{t} - p_{r} \propto 
(\frac{r}{R})^{n} \exp{-(\frac{r}{R})},$ with  $n=1,2,3,4$.

\setlength{\baselineskip}{8.7mm}

{\section{Exact Solutions For  $ \rho \propto \srr$}}

\noindent We will now consider anisotropic stellar configurations with
the following
expression for the energy density
\be
\label{density}
\rho = \frac{1}{8 \pi} \left( \frac{a}{r^{2}} + 3b \right),
\ee
\noindent where both  $a$ and $b$ are constant. The choice of the values for 
$a$ and $b$ is dictated by the physical configuration under consideration.
For example, $ a = 3/7$ and $b = 0$, corresponds to a relativistic Fermi gas.
If we take $ a= 3/7$ and $b \neq 0$ then we have a  relativistic Fermi gas
core immersed in a constant density background. For large $r$ the constant
density term dominates ($r_c^2\gg a/3b$), 
and can be thought of as modeling a shell 
surrounding the core. 

In this section we will consider cases where the pressure anisotropy closely
resembles the behavior of the energy density. Hence we will take
\be
\label{ansio}
p_{t}  - p_{r} = \frac{1}{8 \pi}\left( \frac{c}{r^{2}} + d \right)~,
\ee
with $c$ and $d$ constant.
The motivation for these ansatze comes from similar approaches  
used to model the equation
of state of ultradense neutron stars \cite{RUDERMAN,CANUTO}.

We have found it convenient to seek
solutions for the metric function $ \nu(r) $ directly, rather than 
solving the generalized TOV 
equation. We will then use the known functions $ \lambda(r) $ and $ \nu(r) $ 
to find the radial and tangential pressures. From eqs. \ref{einstein.1},
\ref{einstein.2}, and \ref{einstein.3}, 
we find 
\be
\label{inveq}
\left( \frac{{\nu}^{\prime \prime}}{2} + \frac{ ({\nu}^{\prime })^{2}}{4} 
\right)
e^{-\lambda} - {\nu}^{\prime}\left( \frac{{\lambda}^{\prime}}{4} + 
\frac{1}{2r}\right)
e^{-\lambda}  - \left( \frac{1}{r^{2}} + 
\frac{ {\lambda}^{\prime}}{2r}\right)e^{-\lambda} 
+ \frac{1}{r^{2}} = 8 \pi(p_{t} - p_{r}).
\ee
 Introducing  a new variable $y = e^{\frac{\nu}{2}}$, 
eq. \ref{inveq} becomes,
\be
\label{yyy}
( y^{\prime \prime} )
e^{-\lambda} - {y}^{\prime}\left( \frac{{\lambda}^{\prime}}{2} + 
\frac{1}{r}\right)
e^{-\lambda}  -  y \left[( \frac{1}{r^{2}} + 
\frac{ {\lambda}^{\prime}}{2r})e^{-\lambda} 
- \frac{1}{r^{2}} \right] = 8 \pi y(p_{t} - p_{r}). 
\ee
Since $ e^{-\lambda} = 1  - 2m(r)/r$, using eq. \ref{density}
we find
\be
e^{-\lambda} = 1 - a - br^{2} \equiv I_{b}^{2}(r).
\ee
In this section we will define the function$
I_{b}^{2}(x) \equiv 1 - a -bx^{2}$
to simplify our expressions.
When $b = 0$, we will write $I_{0}^{2} \equiv 1 - a$.
Using the expression for $e^{-\lambda}$ in eq. \ref{yyy} and substituting 
for the pressure anisotropy we find
\be
\label{yy2}
\left[br^{4} - (1-a)r^{2} \right]y^{\prime \prime} + (1-a)r y^{\prime} 
-(a -c -dr^{2})y = 0~.
\ee 
We give the full solution of eq. \ref{yy2} with $ a,b,c,d \neq 0$ in 
Appendix B. Below, we will consider solutions obtained from choosing
specific values for $a,b,c$ and $d$.

\noindent
{\bf CASE I:} ~ Stars with no crust ($b = d = 0$) \\ 
 We first consider 
configurations  where the energy density is given by 
\be
\label{invsquare}
 \rho  = \frac{a}{8 \pi r^{2}}. 
\ee
\noindent  For this density profile, the total mass is $ M = aR/2 $ and
\be
\label{lameq}
e^{-\lambda} = 1 - a.
\ee
\noindent  Since for any static spherically-symmetric configuration we expect 
$(2M/R)_{crit} \le 1$,  we must have $ a < 1$. (Also,
 the metric coefficient $g_{rr} $ becomes infinite when $a = 1$).
A density profile  with this  spatial dependence 
on the radial coordinate was  found to be an
exact isotropic solution of the TOV equation for  the interior of 
ultra high-density neutron stars by Misner and Zapolsky \cite{MISNER}.
Assuming that the neutron star core can be modeled as a relativistic Fermi
gas, ${\it i.e.}$,  $ p_{r} = \rho(r)/3 $, they found the  density  to be given 
by eq. \ref{invsquare}, 
with $ a = 3/7$. 
We
note that the Misner-Zapolsky solution cannot be used to construct a complete
star, since this would require the radius of the star to be infinite. Here,
we want to construct stars with finite radii and density given by eq.
\ref{invsquare} in the context of anisotropic pressure. Thus,
we impose boundary conditions such that $p_r(R)=0$. We also note that
Herrera investigated anisotropic solutions with similar energy density, 
in the context of ``cracking,''  when perturbations in the fluid induce 
anisotropic stresses in the star \cite{HERRER,HERRERA}. However, we follow a 
different approach, focusing on the physical properties of static anisotropic
solutions.

\noindent With $ b = d=0$,  eq. \ref{yy2}  reduces to 
 an Euler-Cauchy equation,
\be
(1- a)r^{2} y^{\prime \prime} - (1 - a)r{y}^{\prime}  + (a - c )y =0. 
\ee

The solutions of this equation divide into three classes, depending on the 
value of 
\be
q \equiv  \frac{( 1 +c - 2a)^{\frac{1}{2}}}{(1 - a)^{\frac{1}{2}}}
\ee

\vspace{1.cm}

\noindent {\bf Case I.1}: $q$ is real \\


The solution for $y$ is 
\be
y = A_{+} \left(\frac{r}{R} \right)^{1+q} + 
A_{-} \left(\frac{r}{R} \right)^{1-q}~,
\ee
with the constants $A_{+}$ and $A_{-}$ fixed by boundary conditions. 
For the case under consideration here ($ b = d=0$), the 
boundary conditions are
\be
e^{-\lambda(R)} = e^{\nu(R)}  =  I_{0}^{2}, ~~~{\rm and}
~e^{\nu (R)} \frac{d \nu}{dr}|_R = \frac{a}{R}~.
\ee
 Applying the boundary conditions we find 
\be
A_{+} = \frac{I_{0}}{2} + \frac{1 - 3I_{0}^{2}}{4qI_0} ~~~~~ {\rm and} 
~~~~~A_{-} = A_{+}(q \rightarrow - q).
\ee

\noindent The radial pressure for this case, after substituting the expressions
for $A_{+}$ and $A_{-}$, is 
\be
8 \pi p_{r} = \frac{(3I_{0}^{2} - 1)^{2} - 4q^{2}I_{0}^{4}}{r^{2}}
\left[ \frac{R^{2q} - r^{2q}}{(3I_0^{2} - 1 + 2qI_0^{2})R^{2q}  
+ ( 1 -3I_0^{2} + 2qI_0^{2})r^{2q}} \right]~.
\ee
\noindent We note  that the boundary conditions automatically guarantee that 
$p_r(R)=0$. 
The radial pressure is always greater than zero provided  $ a < 2/3 $ and 
$ a^{2} > 4c(1 - a)$. Since by definition $ a > 0$,
the second condition implies  $c > 0$. Thus, this model does not allow for 
negative 
anisotropy. Further, since we are considering the case $q > 0$, we must impose
the condition $ 1 + c < 2a $.
Combining the two inequalities for $a$ and $c$, we obtain, $ 2a -1 <
c < a^{2}/4(1-a) $. Since we have $ 0 < a < 2/3$ we find that $0 < c < 1/3$.
We note that for the anisotropic case the maximum value of 
$a $ is $ 2/3$, corresponding to a $33 \% $ increase
when compared with the isotropic case ($a=3/7$). 
In figure 4 we plot the radial pressure, $p_{r}$, as a function of 
the radial coordinate $r$, 
for $a = 3/7$ and several values of $c$. Note that for this choice of $a$,
the inequality $c < a^2/4(1-a)$ imposes that $c<0.08$ for positive pressure
solutions. This can be seen in the figure. For larger anisotropies, no static 
self-gravitating stable configuration is possible.

\hspace{2.0in}
\psfig{figure=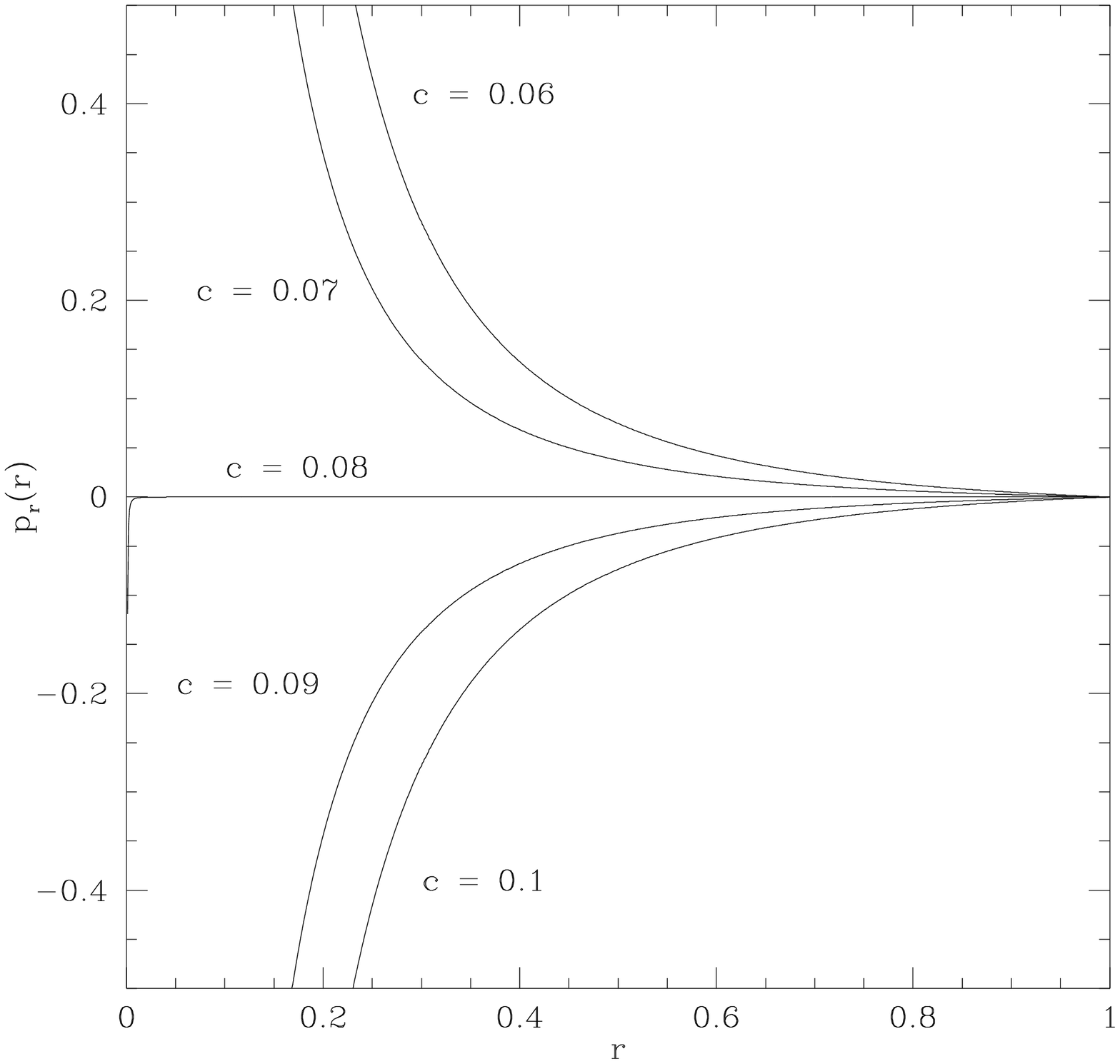,width=2.7in,height=2.5in}

\setlength{\baselineskip}{5mm}

\noindent Fig. 4:  Radial pressure as a function of $r$ for $\rho$ 
and $(p_{t} - p_{r}) \propto r^{-2}$ and $q $ real.

\setlength{\baselineskip}{8.25mm}

For $ r \ll R$,  we find
\be
8 \pi r^2 p_{r} = 3I_0^{2} - 1 - 2qI_0^{2}~.
\ee 

\noindent Choosing $a=3/7$ we recover, in the limit $c\rightarrow 0$, the 
Misner-Zapolsky solution \cite{MISNER}, with 
$ p_{r} = 1/(56 \pi r^{2})=\rho/3$.

\vspace{1.cm}

\noindent {\bf Case I.2: $q = 0$}

\noindent The solution for $y$  with $q = 0$ is  
\be
\label{sola}
y(r) = A_s \left(\frac{r}{R} \right)  + 
B_s  \left( \frac{r}{R} \right)\ln \left( \frac{r}{R} \right)~,
\ee
\noindent where the constants $A_s$ and $B_s$ are determined from
the boundary conditions. We find
\be
A_s = I_0  ~~~~ {\rm and } ~~~~  B_s = \frac{1 - 3I_0^{2}}{2I_0}.
\ee
\noindent The radial pressure is given by
\be
8 \pi r^2 p_{r} = 3I_0^{2} - 1 + \frac{ ( 1 - 3I_0^{2})}
{ 1 + \frac{(1 - 3I_0^{2})}{2I_0^{2}} \ln \left( \frac{r}{R} \right)} .
\ee

The radial pressure is positive provided $ a < 2/3 $. Since  we are 
considering the case $ q = 0$, we must require $c = 2a - 1 $ and
thus $ -1 < c < 1/3$.

\vspace{1.cm}

\noindent {\bf Case I.3:} $q$ is imaginary 

\noindent  The solution for $y(r)$ with $q$ imaginary is 
\be
y = \frac{r}{R} \left \{ A_{s} \cos \left[u \ln \left 
( \frac{r}{R} \right) \right]
 +  B_{s} \sin \left[u \ln \left( \frac{r}{R} \right) \right] \right \}
\ee
\noindent with $ u = |q|$.
For $A_s$ and $B_s$ we find 
\be
A_s = I_0 ,~~~~~{\rm and} ~~~~~ B_s = \frac{1 - 3I_0^{2}}{2uI_0}~.
\ee
\noindent The radial pressure is given by
\be
\label{eqqqq}
8 \pi r^2p_{r} = 3I_0^{2} - 1  +  \frac{2uI_0^{2}
\left[1 - 3I_0^{2} -  2uI_0^{2}\tan \left(u \ln \left(
 \frac{r}{R} \right) \right)\right]} 
{(1 - 3I_0^{2})\tan \left(u \ln \left( \frac{r}{R} \right) \right) 
+ 2uI_0^{2}} ~.
\ee

An analysis of eq.  \ref{eqqqq} shows that  $p_{r} > 0 $ provided
that $a < 2/3 $ for any $c$. Also, since we are considering the case
$q$ imaginary here, we require that $ c < 2a - 1$. 

\vfill\eject

\noindent {\bf Case II:} Including a Crust ($a,b,c \neq 0, d = 0$) \\

We will now derive solutions of the field equations 
with the  density profile given by eq. \ref{density}.
This density profile is  essentially a combination of the two
profiles ($ \rho = const$ and  $\rho \propto 1/r^{2}$) that we have studied 
so far. We may think of this situation as modelling an ultradense core immersed
in a background of constant density; at large distances from the core 
($r_c^2 \gg a/3b$), the 
constant density shell dominates the energy density. As in Case I, 
we impose boundary
conditions such that $p_r(R)=0$.

For this density profile, 
\be
e^{-\lambda} = I_{b}^{2}(r)~,
\ee
\noindent  and the equation for $ y = e^{ \frac{\nu}{2}}$ is
\be
\left( (1-a)r^{2}  - br^{4} \right)y^{\prime \prime} -(1 -a)r y^{\prime}
+ ay = 8 \pi r^{2}y(p_{t} - p_{r}).
\ee
The boundary conditions are
\be
\label{bbc}
y(R) = I_{b}^{2} (R) ~~~{\rm and}~~~
y^{\prime}(R) = \frac{ 1 - {I_{b}^{2}(R)}}{2RI_{b}(R)}.
\ee
\noindent Here, we have chosen the ansatz for the anisotropic pressure to be
the same as the case with $b = 0$, that is, we do not include a constant
contribution to the anisotropy ($d=0$ in eq. 53). The general solution for
$d\neq 0$ is given in Appendix B. 
As with the case above with $ b = 0$, there are three classes of 
solutions depending on the value of 
\be
q = {\frac{ ( 1 - 2a + c)}{( 1 - a)}}^{\frac{1}{2}}.
\ee

\vspace{1.cm}
\noindent {\bf Case II.1}   $ q$ real

For $q$ real ($1+c > 2a$) the solution is  
\be
 y = 
A_{+} \left({\frac{r}{R}} \right)^{1 +q}\left[ \frac{ I_{0} + I_b(r)}
{I_0 + I_b(R)}\right]^{-q} +
A_{-} \left({\frac{r}{R}} \right)^{1 -q}\left[ \frac{ I_0 + I_b(r)}
{I_0 + I_b(R)}\right]^{+q}~. 
\ee
Using the boundary conditions as before we find
\be
A_{+} = \left[ \frac{( 3I_{b}^{2}(R) - 1)}{4q I_{0}} + \frac{I_b(R)}{2}
 \right] ~~~~ {\rm and} ~~~
A_{-} = A_{+}( q \rightarrow -q).
\ee

The radial pressure is given by
\be
8\pi r^2p_{r} =   
\frac{\left [3I_{b}^{2}(r) -1 - 2q I_0 I_{b}(r)\right ] + 
\left [3I_{b}^{2}(r) -1 + 2q I_0 I_{b}(r)\right ]
\left(\frac{A_{-}}{A_{+}}\right) 
\left( \frac{r}{R} \right)^{-2q} 
\left( \frac{I_0 + I_{b}(r)}{I_0  + I_{b}(R)} \right)^{2q} }
{1 + \left(\frac{A_{-}}{A_{+}}\right) 
\left( \frac{r}{R} \right)^{-2q} 
\left(\frac{I_{0} + I_b(r)}{I_0 + I_b(R)} \right)^{2q}}. 
\ee

We note that when $b = 0$, we recover the solution with 
no crust. Also, if  we take $a = 0$ and $ b \neq 0$, we have a 
new class of anisotropic solutions with constant density, and anisotropy
proportional to $r^{-2}$.

\vspace{1cm}
\noindent {\bf Case II.2: $q = 0$}

When  $q = 0$, the solution is
\be
y(r) = e^{\nu(r)/2}
= A_{s} \left(\frac{r}{R} \right) + B_{s}\left(\frac{r}{R} \right) 
\ln \left \{ \frac{[I_0 + I_b(r)] 
[I_0 - I_b(R)]}{[I_0  - I_b(r)]
[I_0 + I_b(R)]} \right \} 
\ee
\noindent with
\be
A_{s} = I_b(R),~{\rm and}~ B_{s} = \frac{(3I_{b}^{2}(R) - 1)}{4 I_0 }~. 
\ee
Here, the radial pressure is given by
\be
8 \pi r^2p_{r} = 3I_{b}^{2}(r)^{2} - 1 - 
4 I_b(r) I_0 B_{s} e^{-\frac{\nu}{2}} ~.
\ee

\vspace{1cm}
\noindent {\bf Case II.3:} q is imaginary \\

When  $q$ is imaginary the solution is 
\be
y =
  \left(\frac{r}{R}\right)\left[ 
 A_s \frac{\sin \left(s \ln F(r) \right)}{\sin\left(s \ln F(R) \right)} +
 B_s\frac{\cos \left(s \ln F(r) \right)}{\cos\left(s \ln F(R) \right)} 
\right]~,
\ee
\noindent with
\be
A_s = 2I_{b}(R) \tan(F(R)) \tan(2F(R)) - \frac{(1 - 3I_{b}^{2}(R))}
{s I_0} \tan(2F), ~~
B_s = I_{b}(R) - A_s~,
\ee
\noindent and
\be
 F(x) = s \ln \left[ \frac{(1-a)^{\frac{1}{2}} +
 ( 1 - a -bx^{2})^{\frac{1}{2}}}{(1-a)^{\frac{1}{2}}} \right] ~.
\ee

We intend to investigate numerically the allowed range of the parameters
$a,~b,~c,$ and $d$, as
well as the stability of these solutions, in a forthcoming publication.

\vspace{0.5in}

\section{Conclusions}

We have presented two broad classes of general relativistic
exact solutions of spherically symmetric
stellar configurations exhibiting anisotropic pressure. Our motivation was to
explore the changes in the general properties of the stars induced by varying
amounts of anisotropy. In particular, to each of the solutions we have
demonstrated that anisotropy may indeed change the critical mass and
surface redshift of the equilibrium
configurations, results which we believe are of interest to the astrophysical
community and will stimulate further investigation.
It is an open question if, indeed, anisotropy is revelant for
compact objects, as, for example, ultradense neutron stars. We have 
motivated our results based on past work on this subject, where isotropic
equations of state are at best a reasonable hypothesis. Given that we do not
have as of yet a complete understanding of the physical processes controlled
by strong interactions in ultradense matter \cite{GLENDENNING}, 
it is wise to keep an open
mind to the possibility that anisotropic stresses
do occur. Furthermore, hypothetical
compact objects made of gravitational bound states 
of bosonic fields, the so-called boson stars, are naturally
anisotropic \cite{GLEISER}. 
These objects have been extensively studied in the literature,
as they represent an interesting new class of compact objects whose stability
against gravitational collapse
comes from a combination of Heisenberg's uncertainty principle and 
model-dependent
self-interactions. They may also have some connection with dark matter, if
perturbations on a fundamental scalar field lead to instabilities which 
trigger the 
gravitational collapse of overdense regions.

We have divided our work into two classes of solutions, those with a constant
energy density and those with an energy density falling as $r^{-2}$. Within
each of these classes we presented several possible cases, which we hope
approximate to some extent possible realistic objects, including a 
combined situation where the star's energy density has an ultradense interior
($\rho \propto r^{-2}$) immersed in a shell of constant density. We intend to
study the stability of these solutions against radial perturbations, as well
as provide a detailed numerical analysis of the allowed parameter 
space, in a future work. We also added two appendices, 
one proving the equality between the Tolman and the 
Schwarzschild mass formulas and the other providing 
the general solution for the $\rho \propto r^{-2}+ {\rm const}$ 
case in terms of hypergeometric
functions.

\section{Acknowledgements}

We would like to thank Joseph Harris, Paul Haines, and especially Vincent
Moncrief for their interest in our work and many useful suggestions. 
KD thanks Dartmouth College for a Dartmouth Fellowship. MG was
supported in part by National
Science Foundation Grants PHY-0070554 and PHYS-9453431.  

\section{Appendix A}
Here we will establish the equivalence of the Tolman and Schwarzschild mass
formulas for static spherically symmetric spacetimes with anisotropic 
pressure. 
The  general expression for the Tolman mass formula is 
\be
\label{tov1}
M_{T} = \int \left( 2T^{0}_{0} - T^{\nu}_{\nu} \right) 
\left(-g \right)^{\frac{1}{2}}d^{3}x~.
\ee
Here we have 
\be 
T^{0}_{0} = \rho   \hspace{0.5in} {\rm and} \hspace{0.5in} 
T^{\nu}_{\nu} = \rho - p_{r} - 2p_{t}~.
\ee
\noindent Substituting the values of $T^{0}_{0}$ and $T^{\nu}_{\nu}$ in 
equation \ref{tov1}
we  have
\be
M_{T} = \int_{0}^{R} 4 \pi r^{2} e^{ \frac{\nu + \lambda}{2}}
( \rho + p_{r} + 2p_{t})dr~.
\ee
\noindent Let
\be
I_{1}= \int_{0}^{R} 4 \pi r^{2} \rho e^{ \frac{\nu + \lambda}{2}} dr~{\rm and}~
I_{2}= \int_{0}^{R} 4 \pi r^{2} ( p_{r} + 2p_{t})
 e^{ \frac{ \lambda + \nu}{2}} dr~;
\ee
\noindent then,
\be
M_{T} = I_{1} + I_{2}.
\ee
\noindent Defining
\be
m(r) = \int_{0}^{r} 4 \pi r^{2}\rho dr~,
\ee
\noindent the Schwarzschild mass is given by 
\be
M_{S} = m(R)~.
\ee

\noindent Consider the integral $I_{1}$. Performing an integration by parts
 we find  
\be
\label{tov2}
I_{1} = \left[m e^{ \frac{ \lambda + \nu}{2}} \right]_{0}^{R} -
\int_{0}^{R} m \left(e^{ \frac{ \lambda + \nu}{2}} \right)^{\prime} dr.
\ee
The first term evaluates to $M_{S}$, since $m(0) = 0$ and, at $ r = R, 
\lambda + \nu = 0$. From the field equations it follows that
\be
\label{tov3}
\frac{( \lambda + \nu)^{\prime}}{2} = 
\frac{ 4 \pi r^{2}(\rho + p_{r})}{ (r  - 2m)}. 
\ee
\noindent Substituting
 this expression in equation \ref{tov2} we find 
\be
I_{1} = M_{S} -
\int_{0}^{R}   \frac{ 4 \pi  mr^{2}(\rho + p_{r})}{ (r  - 2m)}
e^{ \frac{ \lambda + \nu}{2}} dr.
\ee
\noindent  Next we consider the integral $I_{2}$. Using the generalized TOV
equation we can substitute for $p_{t}$ and  write
\be
I_{2} = \int_{0}^{R}  4 \pi r^{2} e^{\lambda + \nu}\left( 3p_{r}
 + r \frac{dp_{r}}{dr} +\frac{(4 \pi p_{r} r^{3} + m)(\rho + p_{r})}
{ (r - 2m)} \right)dr~.
\ee
\noindent We will rewrite this as  
\be 
I_{2} = I_{2a} + I_{2b}~,
\ee
\noindent with
\be
I_{2a} =  \int_{0}^{R}4\pi r^{2} (3 p_{r})e^{\frac{{\lambda + \nu}}{2}}dr~,
\ee
\noindent and
\be
I_{2b} =  \int_{0}^{R} 4 \pi r^2 \left( r \frac{dp_{r}}{dr} 
+\frac{(4 \pi p_{r} r^{3} + m)(\rho + p_{r})}{ (r - 2m)} 
\right)e^{\frac{{\lambda + \nu}}{2}}dr~.
\ee
\noindent Integrating $I_{2a}$ by parts we find 
\be
I_{2a} = \left[ 4 \pi r^{3} p_{r}e^{\frac{{\lambda + \nu}}{2}} 
\right]_{0}^{R}  -
\int_{0}^{R} 4 \pi r^{3} \left( \frac{dp_{r}}{dr} + 
\frac{p \left( {\lambda}^{\prime}+ 
{\nu}^{\prime} \right)}{2} \right)e^{\frac{{\lambda + \nu}}{2}}dr~.
\ee
\noindent The first term of this equation  is equal to zero, 
since $p_{r}(R) =0$,
and substituting for $ {\nu}^{\prime} + {\lambda}^{\prime} $ from equation
 \ref{tov3}  we find that 
\be
I_{2a} =  - \int_{0}^{R} 4 \pi r^{2} \left( r\frac{dp_{r}}{dr} + 
 \frac{ 4 \pi p_{r} r^{3}(\rho + p_{r})}{ (r  - 2m)} 
\right)
e^{\frac{{\lambda + \nu}}{2}}dr~.
\ee
\noindent By adding  $ I_{2a} $ and $I_{2b}$, we get
\be
I_{2} =  \int_{0}^{R} 4 \pi r^2 \left ( \frac{m(\rho + p_{r})}{ (r - 2m)} 
\right)
e^{\frac{{\lambda + \nu}}{2}}dr.
\ee
Finally  $I_{1}$ plus $I_{2}$  gives 
\be
M_{T} = M_{S}~.
\ee
\noindent This demonstrates the equivalence of the Tolman and Schwarzschild
mass formulas for static spherically symmetric spacetimes with anisotropic
pressures.

\section{Appendix B}
The  solution for eq. \ref{yy2}  with  $ a,b,c,d \neq  0$, $ a\neq 1$
and  boundary conditions given by eq. \ref{bbc} is
\be 
y = A_{+} \left( \frac{r}{R} \right)^{1 + q} F_{+}(\tilde{r}) 
+  A_{-} \left( \frac{r}{R} \right)^{1 - q} F_{-}(\tilde{r}) 
\ee
Here  $\tilde{r} = \frac{br}{1-a}$ and  $F$ is the hypergeometric function:
\be
 F = F(\alpha,\beta,\gamma,x) = 1 + \sum_{k=1}^{\infty} 
\frac{ (\alpha)_k (\beta)_k}
{(\gamma)_k}  \frac{x^{k}}{k }, ~~~~
(\alpha)_k = \alpha (\alpha +1) ...(\alpha +k -1)~,
\ee
\be
\frac{d}{dx}F(\alpha, \beta, \gamma, x) = F^{\prime} = \frac{ \alpha \beta}
{\gamma}F( \alpha +1, \beta +1, \gamma +1 ,x)
\ee
\noindent and
\be
F_{+} = F( \alpha_+, \beta_+ , \gamma_+ , \tilde{r}), ~~~~
F_- = F_+( q \rightarrow -q)
\ee 
\noindent with
\bea
\alpha_+ & = & \frac{1}{4} [ 1 + 2q - (\frac{b - 4d}{b} ) ]~,  \nonumber \\
\beta_+  & =  & \frac{1}{4} [ 1 + 2q +  (\frac{b - 4d}{b}) ]~,  \nonumber  \\
\gamma_+ &=  & \frac{1}{4}(1 + q)~, \nonumber  \\
q &  =  &\left( \frac{1 - 2a + c}{1 - a} \right)^{\frac{1}{2}}~.
\eea
Using the boundary conditions we find
\be
A_+ = \frac{ 1 - 3I_b^{2}(R) - 2qI_b^{2}(R) + 2I_b^{2}(R) 
\{\ln[F_{-}(\tilde{R})] \}
^{\prime} }
{2I_b(R) \{ qF_{+}(\tilde{R}) + R[F_{+}(\tilde{R})]^{\prime} - RF_{+}(\tilde{R})
\{\ln[F_{-}(\tilde{R})] \} \}^{\prime} }, ~{\rm and}  
~A_{-} = A_{+}(q \rightarrow -q).
\ee
The radial pressure is given by
\bea 
8 \pi r^2p_{r} &= &  \frac{   3I_b^{2}(r) - 1 + 2qI_b^{2}(r) + 2I_b^{2}(r) 
(\ln[F_{+}
(\tilde{r})])^{\prime}}{ 1   + 
\frac{A_{-}}{A_{+}}\frac{F_{-}(\tilde{r})}{F_{+}(\tilde{r})}
(\frac{r}{R})^{-2q}} 
 \\ \nonumber
&+& \frac{(3I_b^{2}(r) - 1 - 2qI_b^{2}(r))
\frac{A_{-}}{A_{+}} (\frac{r}{R})^{-2q}(
\frac{F_{-}(\tilde{r})}{F_{+}(\tilde{r})}
 + 2g^{2}\frac{(F_{-}(\tilde{r}))^{\prime}}
{F_{+}(\tilde{r})})}
{ 1   + \frac{A_{-}}{A_{+}}\frac{F_{-}(\tilde{r})}{F_{+}(\tilde{r})}
(\frac{r}{R})^{-2q}}~. 
\eea

\end{document}